\documentclass[a4paper, amsfonts, amssymb, amsmath, reprint, showkeys, nofootinbib, onecolumn]{revtex4-1}
\usepackage[english]{babel}
\usepackage{graphicx}
\usepackage{dcolumn}
\usepackage{bm}
\usepackage{subcaption}
\usepackage{comment}
\usepackage{xcolor}
\usepackage{float}
\usepackage{titlesec}
\usepackage[pdftex, pdftitle={Article}, pdfauthor={Author}]{hyperref}
\definecolor{purple}{rgb}{0.58,0.0,0.83}

\definecolor{blue(pigment)}{rgb}{0.2, 0.2, 0.6}

\usepackage{scalerel}
\usepackage{tikz}
\usetikzlibrary{svg.path}
\usepackage[normalem]{ulem}

\definecolor{orcidlogocol}{HTML}{A6CE39}
\tikzset{
  orcidlogo/.pic={
    \fill[orcidlogocol] svg{M256,128c0,70.7-57.3,128-128,128C57.3,256,0,198.7,0,128C0,57.3,57.3,0,128,0C198.7,0,256,57.3,256,128z};
    \fill[white] svg{M86.3,186.2H70.9V79.1h15.4v48.4V186.2z}
                 svg{M108.9,79.1h41.6c39.6,0,57,28.3,57,53.6c0,27.5-21.5,53.6-56.8,53.6h-41.8V79.1z M124.3,172.4h24.5c34.9,0,42.9-26.5,42.9-39.7c0-21.5-13.7-39.7-43.7-39.7h-23.7V172.4z}
                 svg{M88.7,56.8c0,5.5-4.5,10.1-10.1,10.1c-5.6,0-10.1-4.6-10.1-10.1c0-5.6,4.5-10.1,10.1-10.1C84.2,46.7,88.7,51.3,88.7,56.8z};
  }
}

\newcommand\orcidicon[1]{\href{https://orcid.org/#1}{\mbox{\scalerel*{
\begin{tikzpicture}[yscale=-1,transform shape]
\pic{orcidlogo};
\end{tikzpicture}
}{|}}}}

\begin{document}

\title{Thermodynamic behavior of cosmological models with fractional entropy}

\author{Miguel Cruz$^1$\orcidicon{0000-0003-3826-1321}}
\email{miguelcruz02@uv.mx}

\author{Diego da Silva$^2$ \orcidicon{0009-0002-7243-9330}}
\email{diego.dasilva.c@mail.pucv.cl}

\author{Simón González$^2$\orcidicon{0009-0005-3488-8751}}
\email{matias.gonzalez.l01@mail.pucv.cl}
\author{Samuel Lepe$^2$\orcidicon{0000-0002-3464-8337}}
\email{samuel.lepe@pucv.cl}

\author{Joel Saavedra$^2$\orcidicon{0000-0002-1430-3008}}
\email{joel.saavedra@pucv.cl}

\author{Manuel Gonzalez-Espinoza$^3$\orcidicon{0000-0003-0961-8029}}
\email{manuel.gonzalez@upla.cl}

\affiliation{$^1$Facultad de F\'{\i}sica, Universidad Veracruzana 91097, Xalapa, Veracruz, M\'exico.\\
$^2$Instituto de F\'\i sica, Pontificia Universidad Cat\'olica de Valpara\'\i so, Casilla 4950, Valpara\'\i so, Chile.\\
$^3$Laboratorio de investigaci\'on de C\'omputo de F\'isica, Facultad de Ciencias Naturales y Exactas, Universidad de Playa Ancha, Subida Leopoldo Carvallo 270, Valpara\'iso, Chile.}

\date{\today}

\begin{abstract}
We investigate the thermodynamic and phenomenological implications of a cosmological model governed by fractional entropy applied to the apparent horizon of a flat Friedmann-Lema\^{i}tre-Robertson-Walker (FLRW) universe. By utilizing the unified first law of thermodynamics alongside the Kodama-Hayward temperature, we derive a generalized set of Friedmann equations characterized by a fractional parameter $\alpha \in (1,2]$. The thermodynamic analysis reveals that the specific heats $C_V$ and $C_p$ share the same sign and depend solely on the deceleration parameter, demonstrating that the fractional model is thermodynamically stable during the late-time accelerated expansion and does not exhibit phase transitions. To constrain the background dynamics, we confront the truncated fractional model with a joint sample of late-time observational data, including Cosmic Chronometers, Pantheon+SH0ES supernovae, and the latest DESI DR2 Baryon Acoustic Oscillations. Exploring the physically motivated range $1<\alpha\le 2$, we find that the fit quality degrades monotonically as $\alpha$ decreases from the General Relativity limit. Rather than limiting the model's physical value, this demonstrates its theoretical robustness: the fractional framework acts as a continuous deformation parameter that preserves macroscopic thermodynamic stability. The data favors $\alpha$ close to 2 (yielding $H_0=69.50\pm 0.42$ km/s/Mpc and $\Omega_{m0}=0.292\pm 0.008$), revealing that while the late-time background expansion strongly constrains deviations from the standard area law, the fractional model smoothly and stably accommodates these constraints without exhibiting thermodynamic pathologies.
\end{abstract}


\maketitle

\section{Introduction}

The realization that the dynamics of spacetime can be cast in thermodynamic language has profoundly altered our view of gravity. The seminal contributions of Bekenstein, Hawking, and Gibbons \cite{Bekenstein:1973ur,Bardeen:1973gs,Hawking:1975vcx,Gibbons:1976ue} demonstrated that black holes behave as thermodynamic objects, possessing an entropy proportional to the area of their event horizon, $S = A/4$, and a temperature fixed by the surface gravity; see also, for example, Refs. \cite{Kodama:1979vn, Hayward:1993wb, Hayward:1994bu}. This entropy–area law, universal within Einstein’s theory of general relativity, suggests that the fundamental degrees of freedom of spacetime scale with surface area rather than volume, providing deep insights into the holographic nature of gravity. Soon thereafter, Jacobson demonstrated that Einstein’s equations themselves can be derived from the Clausius relation, $\delta Q = TdS$, applied to local Rindler horizons \cite{Jacobson:1995ab}. These insights firmly establish the paradigm that gravitational field equations and thermodynamic relations are two complementary descriptions of the same underlying physics.

The universality of the entropy–area law, however, is expected to break down when quantum or statistical corrections are taken into account. In particular, microscopic state-counting arguments, statistical fluctuations, and effective theories of quantum gravity generically lead to modifications of the entropy functional. Several proposals have been introduced, including logarithmic and power-law corrections inspired by loop quantum gravity \cite{Zhang:2008gt}, Tsallis and Kaniadakis entropies based on non-extensive statistical mechanics \cite{Kaniadakis:2002zz, Kaniadakis:2005zk}, and Barrow entropy, motivated by fractal deformations of the horizon surface \cite{Barrow:2020tzx}. Each of these generalized entropies leads to modified gravitational dynamics when applied to the apparent horizon of an FLRW universe, producing corrections to the Friedmann equations and altering cosmic expansion \cite{Sheykhi:2010zz,Sheykhi:2021fwh,Sheykhi:2023aqa}. This thermodynamic approach, therefore, offers a powerful and unifying framework: by choosing an entropy functional, one effectively selects the gravitational theory governing cosmic evolution.

Within this broad class of generalizations, fractional entropy stands out as a particularly natural and mathematically rich extension. The idea is to generalize the scaling of entropy with area using concepts from fractional calculus and anomalous geometry. A common definition of fractional entropy is
\begin{equation}
S_{\alpha} = \gamma \, A^{\alpha},
\end{equation}
where $0<\alpha \leq 1$ is the fractional index that characterizes the non-additivity properties of the entropy, $\gamma$ is a normalization constant, and $A$ is the horizon area. For $\alpha = 1$, the standard Bekenstein–Hawking entropy is recovered, while deviations $\alpha \neq 1$ encode new scaling regimes that may arise if spacetime has a fractal-like microstructure or if the underlying statistical mechanics deviates from the Boltzmann–Gibbs paradigm. Unlike additive corrections, which typically appear as subleading terms to the area law, the fractional entropy modifies the scaling itself, fundamentally altering the thermodynamic description of horizons. Such a modification is expected to be most relevant in the early universe, where the apparent horizon is small and quantum gravitational effects cannot be neglected. However, it may also leave detectable imprints at late times. This theoretical formulation is not merely phenomenological; it is deeply rooted in recent advances in fractional quantum mechanics and quantum cosmology \cite{Moniz2020, Canedo2025}. Specifically, the use of fractional calculus in analyzing the Wheeler–DeWitt equation and scalar field dynamics has demonstrated that the event horizon of a nucleated universe may possess a fractal structure \cite{Jalalzadeh2022, Rasouli2024}. Moreover, fractional gravity models endowed with nonlocal d'Alembertian operators have provided exact bouncing and de Sitter solutions, further validating the use of fractional scaling in high-energy cosmological regimes \cite{SalvadorGarcia2026, MicoltaRiascos2023}.

Taking fractional entropy into account within cosmological theory leads to a number of profound and wide-ranging consequences. From a thermodynamic perspective, replacing the Bekenstein–Hawking entropy with $S_{\alpha}$ modifies the Clausius relation and the unified first law \cite{Hayward:1997jp,Cai:2005ra, Cai:2006pa, Cai:2006rs}, resulting in corrected Friedmann equations. The dynamics of the universe then depend explicitly on the parameter $\alpha$, which controls the effective gravitational coupling and the relationship between energy density, pressure, and the thermodynamics of the horizon. From the stability perspective, fractional scaling influences the behavior of the heat capacity of the apparent horizon, determining whether the system is thermodynamically stable or unstable. Divergences in the heat capacity indicate the presence of critical points, suggesting the existence of horizon phase transitions analogous to those extensively studied in black hole thermodynamics. Moreover, the Gibbs free energy associated with the apparent horizon may exhibit swallow-tail structures, indicating first-order transitions and the coexistence of cosmological phases. These phenomena reveal a thermodynamic richness in the cosmological horizon that goes well beyond the standard $\Lambda$CDM picture.

From the cosmological perspective, fractional entropy corrections propagate into the expansion history $H(z)$. Since the modifications scale with the horizon area, their effects are most pronounced at high redshift but can also affect the late universe if $\alpha$ deviates significantly from unity. This opens a phenomenological window for testing fractional entropy models against observational datasets, such as Type Ia supernovae (Pantheon+), baryon acoustic oscillations, and cosmic chronometers. In particular, the modification of $H(z)$ provides an avenue to explore whether fractional entropy can ease or resolve the long-standing $H_{0}$ tension between local and early-universe determinations of the Hubble constant. Positive or negative deviations of $\alpha$ relative to unity can suppress or enhance the expansion rate, respectively, thereby shifting the inferred value of $H_{0}$. Current observational constraints can therefore be used to place limits on the fractional parameter and evaluate the viability of the model.

The purpose of this work is to systematically investigate the thermodynamic and cosmological consequences of adopting fractional entropy in an FLRW background. We first derive the modified Friedmann equations from the unified first law of thermodynamics when the entropy takes the fractional form $S_{\alpha}$. We then analyze the effective equation of state associated with the apparent horizon, with special emphasis on heat capacity and thermodynamic stability. Crucially, as we will demonstrate, this fractional framework strictly preserves the macroscopic thermodynamic stability of the late universe; it naturally avoids the pathological horizon phase transitions and future singularities often present in alternative dark energy scenarios, ensuring a continuous, well-behaved expansion history. Finally, we explore the cosmological dynamics implied by the fractional entropy, focusing on observational constraints from late-time probes. Throughout this work, we adopt natural units where $G=c=1=M_{\rm{pl}}$.

\section{Modified Friedmann equations from fractional entropy: full Kodama-Hayward temperature}

First, we would like to discuss the geometric setup and the apparent horizon. Then, we consider an FLRW spacetime with spatial curvature $k$. 
\begin{equation}
ds^{2}=-dt^{2}+a^{2}(t)\!\left(\frac{dr^{2}}{1-kr^{2}}+r^{2}d\Omega^{2}\right),
\qquad
H\equiv \frac{\dot a}{a},\qquad
R_{A}=\frac{1}{\sqrt{H^{2}+k/a^{2}}},
\end{equation}
where the apparent horizon radius $R_{A}$ satisfies the condition $h^{ij}\partial_{i}R\partial_{j}R=0$, with $h_{ij}$ being the metric defined as $h_{ij} := \mbox{diag}[-1,a^{2}(t)/(1-kr^{2})]$, along with $i,j=0,1=t,r$ and $a(t)$ being the scale factor. It is useful to introduce the following variables
\begin{equation}
x \;\equiv\; H^{2}+\frac{k}{a^{2}}=\frac{1}{R_A^{2}},
\qquad
X \;\equiv\; \dot H-\frac{k}{a^{2}},  \label{eq:vars}
\end{equation}
a straightforward calculation yields
\begin{equation}
\frac{\dot R_A}{2HR_A} = -\frac{1}{2}R_A^{2}X.
\label{eq:RAdot}
\end{equation}
Now, we would like to compute the full Kodama-Hayward temperature using the fractional entropy correction to observe the thermodynamic consequences. The Kodama–Hayward surface gravity and the corresponding horizon temperature read
\begin{equation}
\kappa=-\frac{1}{R_A}\!\left(1-\frac{\dot R_A}{2 H R_A}\right),
\qquad
T_h=-\frac{\kappa}{2\pi}
=\frac{1}{2\pi R_A}\!\left(1-\frac{\dot R_A}{2 H R_A}\right).
\label{eq:Tfull}
\end{equation}
The nature of the horizon defines the sign of the temperature associated with the apparent horizon. As discussed in Refs. \cite{horizon1, horizon2, horizon3}, the expanding cosmology corresponds to a past-inner trapping horizon. In this case, the surface gravity satisfies $\kappa < 0$ and $T_{h} \propto -\kappa$. These conditions lead to a positive physical temperature. For the entropy of the apparent horizon, we will consider the specific form of the fractional entropy equation that was originally derived in Ref. \cite{fractional}\footnote{The black hole thermodynamics was studied in the context of fractional quantum mechanics. Some different extensions for the Schrödinger equation can be elaborated. However, as a starting point, the Hamiltonian
\begin{equation*}
    H=\frac{\mathbf{p}^{2}}{2m}+V(\mathbf{r},t),
\end{equation*}
is generalized to \cite{levy}
\begin{equation*}
    H_{\alpha}(\mathbf{P,r}) \equiv D_{a}|\mathbf{p}|^{\alpha}+V(\mathbf{r}), \ \ \ 1 < \alpha \leq 2,
\end{equation*}
where $\alpha$ is known as L\'evy's fractional parameter and is associated with the concept of path and its induced fractal dimension. For instance, in the Feynman path integral, the measure is generated by the process of Brownian motion, and the dimension is $d = 2$. The L\'evy path integral leads to the dimension $d=\alpha$. See also \cite{Jalalzadeh:2024qej, Mureika:2006tz}.} in terms of the area, $A=4\pi R^{2}_{A}$, as follows
\begin{equation}
S_h= \left(\frac{A}{4}\right)^{\frac{2+\alpha}{2\alpha}} + \Theta(\alpha)\left(\frac{A}{4}\right)^{1-\frac{1}{2}\Delta},
\qquad 1<\alpha < 2,
\label{eq:Sfrac1}
\end{equation}
where $\Delta = (2+\alpha)/\alpha$ defines the fractal dimension, and according to the definition of $\alpha$, we have $2 < \Delta < 3$; additionally, 
\begin{equation}
    \Theta(\alpha) \equiv \frac{\alpha(4-\alpha)(2+\alpha)4^{(\alpha+2)/2\alpha}\pi^{(3\alpha+2)/2\alpha}\Gamma(\frac{2}{\alpha})}{(2-\alpha)\left(\Gamma(\frac{1}{\alpha})\right)^{2}}, \qquad \alpha \neq 2, \label{eq:coef}
\end{equation}
where $\Gamma(x)$ is the gamma function. It is important to explicitly state that the foundational derivation of generalized cosmological equations combining the Kodama-Hayward temperature with fractional entropy expressions has been pioneered in recent literature. In particular, the rigorous conceptual mapping from fractional quantum mechanics to the thermodynamic properties of the apparent horizon was established by Jalalzadeh, Moniz, and collaborators \cite{Jalalzadeh2022, Rasouli2024}. Our analysis builds upon these pioneering formalisms to specifically test the late-time thermodynamic stability and observational viability of the truncated expansion. Notice that the entropy increases by $dS_{h}$ as the radius of the horizon $R_{A}$ increases by $dR_{A}$. On the other hand, the case $\alpha=2$ leads to the following form for the entropy
\begin{equation}
S_h= \frac{A}{4} - 2\pi \theta \ln \left (\frac{A}{4}\right),
\label{eq:Sfrac2}
\end{equation}
where we introduce $\theta$ as a tracking parameter in order to establish a comparison between the results and the standard case. So, we can use the unified first law (UFL) and the Clausius relation; the UFL \cite{Hayward:1997jp,Cai:2005ra, Li:2013fop, Sebastiani:2023brr} reads
\begin{equation}
dE=T_h\,dS_h+W\,dV, \label{eq:first}
\end{equation}
where the total amount of energy inside the apparent horizon is simply $E=\rho V$, with $V=\frac{4\pi}{3}R_A^{3}$ being the volume enclosed by the apparent horizon. The work density is defined as the two-dimensional normal trace of the energy-momentum tensor, yielding
\begin{equation}
    W \equiv -\frac{1}{2}h^{ij}T_{ij} = \frac{1}{2}(\rho-p),
\end{equation}
with $T_{ij}$ being the energy-momentum tensor of a perfect fluid with energy density $\rho$ and pressure $p$. The first law given in (\ref{eq:first}) can be written as
\begin{equation}
    T_{h}dS_{h} = -dE +WdV,
\end{equation}
where the minus sign indicates that, due to the expansion of the universe, there is a decrease in the internal energy of matter fields inside the horizon. If we consider that the heat flux is dissipated from the system (horizon) to the surroundings \cite{flux}, then the Clausius relation must be $\delta Q = -T_{h}dS_{h}$. Considering the derivative w.r.t time yields
\begin{equation}
    T_{h}\dot{S}_{h} = -\dot{\rho}V -\frac{1}{2}(\rho+p)\dot{V} = (\rho + p)\left\lbrace 3HV-\frac{1}{2}\dot{V} \right\rbrace = 4\pi(\rho+p)HR^{3}_{A}\left(1+\frac{1}{2}R_{A}^{2}X\right)=8\pi^{2}(\rho+p)HR^{4}_{A}T_{h},
\end{equation}
therefore
\begin{equation}
    \dot{S}_{h} = 8\pi^{2}(\rho+p)HR^{4}_{A}.\label{eq:dot1}
\end{equation}
The horizon entropy is (\ref{eq:Sfrac1})
\begin{equation}
    S_{h} = S^{\Delta/2}_{BH}+\Theta(\alpha)S^{1-\Delta/2}_{BH},
\end{equation}
from which we obtain
\begin{eqnarray}
    \dot{S}_{h} = \frac{\partial S_{h}}{\partial S_{BH}}\dot{S}_{BH} &=& -\Delta \pi HR^{4}_{A}XS^{\frac{1}{2}(\Delta-2)}_{BH} - \Theta(\alpha)\frac{\Delta}{2}(2-\Delta)\pi HR^{4}_{A}XS^{\frac{\Delta}{4}(2-\Delta)-1}_{BH},\nonumber \\ &=& -\Delta \pi^{\Delta/2}HR^{\Delta + 2}_{A}X-\Theta(\alpha)\frac{\Delta}{2}(2-\Delta)\pi^{\frac{\Delta}{4}(2-\Delta)}H R^{\frac{\Delta}{2}(2-\Delta)+2}_{A}X, \label{eq:dot2}
\end{eqnarray}
where we used $\dot{S}_{BH}=-2\pi HR^{4}_{A}X$, equating (\ref{eq:dot1}) and (\ref{eq:dot2}), we obtain the exact acceleration equation
\begin{equation}
    \frac{\Delta \pi^{(\Delta/2)-1}}{2}\frac{X}{x^{(\Delta/2)-1}}+\Theta(\alpha)\frac{\Delta}{4}(2-\Delta)\pi^{\frac{\Delta}{4}(2-\Delta)-1}\frac{X}{x^{\frac{\Delta}{4}(2-\Delta)-1}}=-4\pi(\rho+p).\label{eq:accel}
\end{equation}
On the other hand, using the continuity equation $\dot\rho+3H(\rho+p)=0$ and $\dot x=2HX$ from (\ref{eq:RAdot}), yielding $H=\dot{x}/(2X)$, we can write the acceleration equation (\ref{eq:accel}) as follows:
\begin{equation}
    \frac{\Delta \pi^{(\Delta/2)-1}}{2}\frac{\dot{x}}{x^{(\Delta/2)-1}}+\Theta(\alpha)\frac{\Delta}{4}(2-\Delta)\pi^{\frac{\Delta}{4}(2-\Delta)-1}\frac{\dot{x}}{x^{\frac{\Delta}{4}(2-\Delta)-1}}=\frac{8\pi}{3}\dot{\rho}.
\end{equation}
After a straightforward integration, we obtain the following result for the Friedmann constraint from the previous expression:
\begin{equation}
    \frac{\Delta \pi^{\frac{1}{2}(\Delta-2)}}{4-\Delta}x^{\frac{1}{2}(4-\Delta)} +\Theta(\alpha)\frac{\Delta (2-\Delta)\pi^{\frac{\Delta}{4}(2-\Delta)-1}}{(\Delta-4)(\Delta+2)+16}x^{\frac{1}{4}(\Delta-4)(\Delta+2)+4}=\frac{8\pi}{3}\rho, \label{eq:fried}
\end{equation}
where we set the integration constant equal to zero. The complete Friedmann equations (\ref{eq:accel}) and (\ref{eq:fried}), can be rewritten as follows if we consider the variables given in (\ref{eq:vars}) 
\begin{align} 
    & \frac{\Delta \pi^{\frac{1}{2}(\Delta-2)}}{4-\Delta}\left(H^{2}+\frac{k}{a^{2}}\right)^{\frac{1}{2}(4-\Delta)} +\Theta(\alpha)\frac{\Delta (2-\Delta)\pi^{\frac{\Delta}{4}(2-\Delta)-1}}{(\Delta-4)(\Delta+2)+16}\left(H^{2}+\frac{k}{a^{2}}\right)^{\frac{1}{4}(\Delta-4)(\Delta+2)+4}= \frac{8\pi}{3}\rho, \label{eq:comp1}\\
    & \left\lbrace \frac{\Delta \pi^{\frac{1}{2}(\Delta - 2)}}{2}\left(H^{2}+\frac{k}{a^{2}}\right)^{\frac{1}{2}(2-\Delta)}+\Theta(\alpha)\frac{\Delta}{4}(2-\Delta)\pi^{\frac{\Delta}{4}(2-\Delta)-1}\left(H^{2}+\frac{k}{a^{2}}\right)^{1-\frac{\Delta}{4}(2-\Delta)}\right\rbrace \left(\dot{H}-\frac{k}{a^{2}}\right)=-4\pi(\rho+p).\label{eq:comp2}
\end{align}
It is important to note that, in both of the above equations, only the first term on the left-hand side corresponds to the cosmological model analyzed in \cite{Coker_2023}. We will refer to this situation as {\it truncated fractional cosmology}, which occurs when $\Theta(\alpha)=0$. In this framework, we need only set $\Delta =2$ in order to recover the standard cosmological model.\\

If we repeat the procedure outlined previously, but now consider the form of the entropy given in Eq. (\ref{eq:Sfrac2}), the Friedmann equations are 
\begin{align} 
    & \left(H^{2}+\frac{k}{a^{2}}\right) - \pi \theta \left(H^{2}+\frac{k}{a^{2}}\right)^{
2}= \frac{8\pi}{3}\rho, \label{eq:comp3}\\
    & \left\lbrace 1- 2\pi \theta \left(H^{2}+\frac{k}{a^{2}}\right)\right\rbrace \left(\dot{H}-\frac{k}{a^{2}}\right)=-4\pi(\rho+p),\label{eq:comp4}
\end{align}
which corresponds to $\alpha =2$. As can be seen, the standard cosmology case is recovered for $\theta=0$ only. On the other hand, the coefficient $\Theta(\alpha)$ given in Eq. (\ref{eq:coef}) can be expressed in terms of the fractal dimension as follows
\begin{equation}
    \Theta(\Delta)= \frac{4^{\frac{\Delta}{2}}\pi^{\frac{1}{2}(\Delta+2)}(2\Delta)\Gamma(\Delta-1)}{(\Delta-1)^{2}(\Delta-2)\left(\Gamma\left(\frac{\Delta-1}{2}\right)\right)^{2}},
\end{equation}
then, using the expression given above, together with $\Gamma(1)=1$ and $\Gamma(1/2)=\sqrt{\pi}$ in the Eqs. (\ref{eq:comp1}) and (\ref{eq:comp2}), we are allowed to compare with Eqs. (\ref{eq:comp3}) and (\ref{eq:comp4}) for $\Delta=2$ since the factor $\Delta-2$ can be removed from the denominator; therefore, we can establish the following interval for the parameter $\theta$: $0 \leq \theta \leq 4/\pi$, which contains the cosmological standard case and the fractional scenario for $\alpha=2$. From now on, we will focus on the dynamical equations (\ref{eq:comp1}) and (\ref{eq:comp2}) with nil spatial curvature since they represent the general case for fractional cosmology.

\subsection{Generalized Second Law}
\label{sec:gsl}
To assess whether the model is thermodynamically viable, one must explicitly take into account the Generalized Second Law (GSL) of thermodynamics. Following the formalism discussed in Ref.~\cite{flux}, the matter fields enclosed by the apparent horizon must be regarded as an open thermodynamic system, which is reflected in the presence of an energy flux through the boundary. The first law of thermodynamics for the matter fields is given by $T_m dS_m = d(\rho V) + p dV - \mu dN$, where $\mu$ is the chemical potential and $dN$ accounts for the number of particles enclosed by the apparent horizon. Combining this with the temporal evolution of the horizon entropy, the GSL is governed by the total entropy variation $T_h \dot{S}_h + T_m \dot{S}_m \geq 0$. Applying the GSL developed in \cite{flux} to our full fractional entropy functional (\ref{eq:Sfrac1}), which is a linear combination of two scale terms $S_h = S_{BH}^{\Delta/2} + \Theta(\alpha) S_{BH}^{1-\Delta/2}$, we obtain the exact analytical expression for the total entropy rate 
\begin{equation}
    T_h \dot{S}_h + T_m \dot{S}_m = \frac{\Delta}{4} S_{BH}^{\frac{\Delta}{2}-1} \left[ \epsilon^2 - \frac{3(\epsilon-1)}{2 - \frac{\Delta}{2}} \right] + \frac{\Theta(\alpha)}{2} \left( 1 - \frac{\Delta}{2} \right) S_{BH}^{-\frac{\Delta}{2}} \left[ \epsilon^2 - \frac{3(\epsilon-1)}{1 + \frac{\Delta}{2}} \right].
\end{equation}
Here, $S_{BH} = A/4$ denotes the usual Bekenstein–Hawking entropy, $\Delta = \frac{2+\alpha}{\alpha}$, and we follow the notation introduced in \cite{flux} for the parameter $\epsilon$, defined by $\epsilon \equiv -\dot{H}/H^{2}$. During the late-time dark energy era, the universe undergoes accelerated expansion, meaning $\epsilon < 1$, which renders $(\epsilon - 1) < 0$. Within the physically motivated domain of our fractional parameter $1 < \alpha \le 2$, the effective exponent is strictly bounded as $1 \le \Delta/2 < 3/2$. This guarantees that the denominators $(2 - \Delta/2)$ and $(1 + \Delta/2)$ are positive, making the expressions inside both square brackets strictly positive. Crucially, during the late-time accelerated expansion, the apparent horizon area becomes large ($S_{BH} \to \infty$). Analyzing the asymptotic scaling of both terms:
\begin{itemize}
    \item The first term behaves as $S_{BH}^{\frac{2-\alpha}{2\alpha}}$. Since $\alpha$ lies in a range that makes this exponent strictly positive, the term increases with $S_{BH}$ and therefore dominates the total entropy rate.
    \item The second term scales as $S_{BH}^{-\frac{\alpha+2}{2\alpha}}$. Since this exponent is strictly negative, the term rapidly decays to zero at late times.
\end{itemize}
Consequently, at late times, the positive first term dominates over the second, effectively eliminating the correction introduced by $\Theta(\alpha)$. This exact calculation ensures that $dS_{tot}/dt > 0$ throughout the dark energy era, thereby strictly fulfilling the GSL of Thermodynamics.

\subsection{Thermodynamic setup}
Notice that, according to Eq. (\ref{eq:first}), $W$ can be related to the thermodynamic pressure, $P$, since it is the conjugate variable of the thermodynamic volume, we write
\begin{equation}
dE \;=\; -T_h\,dS_h - P_{\rm eff}\,dV ,
\end{equation} 
where we have identified
\begin{equation}
P_{\rm eff} = -W = \frac{p-\rho}{2}.
\label{eq:Peff_def}
\end{equation}
If we consider the flat universe $k=0$ and the Friedmann equations (\ref{eq:comp1}) and (\ref{eq:comp2}), then the equation of state can be written in parametric form as follows:
\begin{align}
\label{eq:EoS_parametric1}
& P_{\rm eff}(T_h,V;\alpha) = -\frac{\Delta \pi^{\frac{1}{2}(\Delta-2)-1}}{16}\left(y^{3-\Delta}z+\frac{6}{4-\Delta}y^{4-\Delta} \right) \nonumber \\ & + \Theta(\Delta)\frac{\Delta(\Delta-2)\pi^{\frac{\Delta}{4}(2-\Delta)-2}}{32}\left(y^{3-\frac{\Delta}{2}(\Delta-2)}z+\frac{12}{(\Delta-4)(\Delta+2)+16}y^{\frac{1}{2}(\Delta-4)(\Delta+2)+8} \right),
\qquad
\left\{
\begin{aligned}
&y \equiv \left(\frac{4\pi}{3V}\right)^{1/3}\!=\frac{1}{R_A},\\[2pt]
&z \equiv 4\pi T_h-2y,
\end{aligned}
\right.
\end{align}
it is worth mentioning that the variable $y$ is introduced through the volume enclosed by the apparent horizon and not by means of the specific volume, $v=2R_{A}$, as is usually done. The critical points of the thermodynamic pressure are defined by van der Waals–type conditions
\begin{equation}
\left(\frac{\partial P_{\rm eff}}{\partial V}\right)_{T_h}=0,
\qquad
\left(\frac{\partial^{2} P_{\rm eff}}{\partial V^{2}}\right)_{T_h}=0.
\label{eq:crit_cond}
\end{equation}
Since $y=(4\pi/3V)^{1/3}$, one may equivalently work at fixed $T_h$ with \( (\partial P_{\rm eff}/\partial y)_{T_h}=0 \) and \( (\partial^{2}P_{\rm eff}/\partial y^{2})_{T_h}=0. \) It is convenient to introduce the dimensionless ratio
\begin{equation}
u \;\equiv\; \frac{z}{y}=\frac{4\pi T_h}{y}-2,
\qquad\Rightarrow\qquad
\frac{du}{dy}\Big|_{T_h}=-\frac{u+2}{y}.
\end{equation}
For the second derivative, we differentiate with respect to $y$ while keeping $T_h$ fixed, using $u'(y)=-(u+2)/y$ and $u''(y)=\partial_y[-(u+2)/y]=\tfrac{u+2}{y^{2}}-\tfrac{u'}{y}=\tfrac{2(u+2)}{y^{2}}$. 
Although the extremality conditions formally admit solutions,
these critical points do not correspond to physical phase
transitions. Instead, they arise as isolated mathematical
solutions that lack the thermodynamic structure required
for phase coexistence. The $P$–$V$ diagram does not show phase coexistence, and the evolution of the Gibbs free energy as a function of pressure does not align with the expectations for a phase transition. The same conclusions are reached when one uses the specific volume instead of the volume within the apparent horizon, or when the truncated version of the model is applied. This result indicates that fractional entropy does not
induce genuine thermodynamic criticality at the level of
cosmological horizons. In contrast to other generalized
entropy frameworks, where phase transitions are associated
with instabilities or abrupt changes in the expansion
dynamics, the present model preserves a smooth and
continuous thermodynamic behavior.

\subsection{Specific heats of the cosmological model}
Using the standard definitions of classical thermodynamics, we can write within the single fluid description
\begin{equation}
    C_{V} = \frac{\partial U}{\partial T}, \qquad C_{p} = \frac{\partial h}{\partial T}, 
\end{equation}
both expressions represent the specific heat at constant volume, $V$, and at constant pressure, $p$, respectively; $U$ and $h$ are the internal energy of the system and its enthalpy. $T$ is the temperature of the fluid. If we consider the volume enclosed by the apparent horizon, then we can write $U = \rho V = (4\pi/3) R^{3}_{A}\rho$ and $h=(\rho + p)V=(4\pi/3)(\rho + p)R^{3}_{A}$. Therefore, since all quantities depend on cosmic time, $C_{V}=(dU/dt)(dT/dt)^{-1}$ and $C_{p}=(dh/dt)(dT/dt)^{-1}$, we obtain for the $k=0$ case
\begin{equation}
C_{V}(t) = \Delta(1+q(t))\Psi(t), \qquad C_{p}(t) = \Delta (1+\omega)(1+q(t))\Psi(t),
\end{equation}
where $q(t)$ is the deceleration parameter. We have introduced $\Psi(t) > 0$ as a strictly positive, time dependent auxiliary function that collects the background variables (such as $H(t)$ and $T(t)$) and the fractional parameter $\Delta$. Because the explicit algebraic form of $\Psi(t)$ is quite cumbersome and does not provide further physical insight for the current analysis, it is omitted here for brevity. The crucial point is that both $C_V(t)$ and $C_p(t)$ share this common positive factor, meaning their thermodynamic behavior and stability are strictly dictated by the evolution of the deceleration parameter $q(t)$. In addition, we have considered a barotropic EoS for the energy density and pressure of the fluid, given as $p=\omega \rho$, where $\omega$ is a constant parameter restricted to $-1 \leq \omega \leq 0$, so that it can account for either dark matter–like or dark energy–like behavior. From the single fluid description for cosmic fluids, the evolution equation for the temperature reads \cite{maartens}
\begin{equation}
    \frac{\dot{T}(t)}{T(t)}=-3H\left(\frac{\partial p}{\partial \rho} \right)_{n}=-3H\omega,
\end{equation}
which remains valid for a fluid with conserved energy and a conserved number of particles. For a positive temperature for the fluid and $H(t) > 0$, since we consider an expanding universe consistent with $-1 \leq q(t) \leq 0$, we see that $C_{V}(t)$ and $C_{p}(t)$ have the same sign. This behavior provides a direct thermodynamic diagnostic of
stability. In particular, the absence of divergences or sign
changes in $C_V$ and $C_P$ rules out both first-order and
second-order phase transitions, establishing that the
fractional cosmological model evolves within a stable
thermodynamic regime. No phase transition is observed, as this kind of process is characterized by conditions $C_{p}\rightarrow \infty$ or $C_{V}\rightarrow \infty$ (or a change of sign in any of the specific heats). A dependence on the deceleration parameter in the specific heats was also obtained in \cite{duary}, where the transition from $q>0$ to $q<0$ given by $q=0$ induces a singularity in $C_{V}$ at some positive value of $z$ and $C_{V}<0$ for $z\simeq 0$, indicating that the transition from the decelerated to the accelerated stage in cosmic expansion can be described as a second order phase transition. As pointed out in Ref. \cite{heat}, within the standard cosmological model, the values $C_{p}=0$ and $C_{V} < 0$ characterize the beginning of the dark energy domination epoch. A vanishing $C_{p}$ may be interpreted as a consequence of the existence of matter-dominated and dark energy dominated epochs. A change of sign in the parameter $q$ at the past would be interpreted as a phase transition in the specific heats, characterized by a null $C_{p}$ and $C_{V}<0$.\\ 

In our case, the specific heats of the model do not exhibit any of the behaviors discussed above. Therefore, even though its Friedmann equations have a complicated structure, this model does not interpret the late-time accelerated expansion of the universe as a thermodynamic phase transition, in contrast to other cosmological scenarios. This outcome corroborates the formulation introduced in the previous section.\\ 

From our results, the adiabatic index is $C_{p}/C_{V}=1+\omega >0$, and this corresponds to an accelerated cosmic expansion since both specific heats have the same sign; see, for instance, Ref. \cite{q}, where it is established that if $C_{p}/C_{V} < 0$, this corresponds to a decelerated universe, and $C_{p}/C_{V} > 0$ to an accelerated one. Therefore, within the single fluid description, this cosmological model is viable only as a description of the late-time accelerated expansion of the universe. We note that for $\omega < -1$ the ratio $C_{p}/C_{V}$ becomes negative, which is inconsistent with an expanding universe. Consequently, these results cannot be applied in the phantom regime. In the limit $\omega, q \rightarrow -1$, both specific heats are zero. Cosmologically, the absence of late-time phase transitions carries profound implications for the viability of the fractional entropy model. In many modified gravity frameworks or extended thermodynamic scenarios, horizon phase transitions correlate with thermodynamic instabilities, abrupt shifts in the effective equation of state, or future singularities (such as Big Rip scenarios). The fact that $C_V$ and $C_p$ remain well-behaved and share the same sign throughout the accelerated phase indicates that the fractional deformation of the horizon area constitutes a smooth, continuous modification. It ensures that the cosmic transition from a decelerated, matter-dominated epoch to an accelerated, dark energy-like phase occurs while strictly preserving the macroscopic thermodynamic stability of the universe. Consequently, the fractional parameter $\alpha$ effectively mimics the phenomenological background dynamics of dark energy without introducing unphysical events or thermodynamic pathologies in the late universe. On the basis of these results, the complete and truncated Friedmann equations of motion are thermodynamically equivalent; therefore, from this point onward, we will concentrate on the truncated model to investigate some of its cosmological consequences.

\section{Truncated Fractional Cosmology}
Starting from the generalized constraint obtained by combining the fractional entropy with the Hayward–Kodama temperature, the equations (\ref{eq:comp1}) and (\ref{eq:comp2}) with $\Theta(\alpha)=k=0$ take the following form:
\begin{equation}
    \Delta\pi^{\frac{\Delta}{2}-1}\frac{1}{4-\Delta}H^{4-\Delta}=\frac{8\pi}{3}\rho, \quad \text{with} \quad \Delta\equiv\frac{2+\alpha}{\alpha}.
\end{equation}
We can perform a change of variable to describe the cosmological quantities in terms of the redshift, $z$, using the usual expression $1+z = 1/a(t)$, assuming a present-day scale factor normalized to unity ($a_0 = 1$). Normalizing at present time, given by $z=0$, so that $H(0)=H_{0}$, we obtain the fractional Friedmann equation in terms of the observational density parameters:
\begin{equation}
    \frac{H(z)}{H_{0}}=\left[\sum_{i}\Omega_{i0}(1+z)^{3(1+w_{i})}\right]^{\frac{\alpha}{3\alpha-2}},
    \label{eq:frac_friedmann}
\end{equation}
where $\Omega_{i0}$ represents the present-day energy density parameters defined with respect to the standard General Relativity (GR) critical density. Imposing the condition $H(0)=H_{0}$ at $z=0$ leads to the normalization relation $\sum_{i}\Omega_{i0} = 1$.

\subsection{Limiting and Special Cases}
To understand the background dynamics, it is instructive to evaluate Equation (\ref{eq:frac_friedmann}) under specific cosmological fluid configurations:
\begin{itemize}
    \item \textbf{Matter and Cosmological Constant ($\Lambda$):} For a universe dominated by pressureless dust ($w_{m}=0$) and a cosmological constant ($w_{\Lambda}=-1$), the expansion rate simplifies to:
    \begin{equation}
        \frac{H(z)}{H_{0}}=\left[\Omega_{m0}(1+z)^{3}+\Omega_{\Lambda }\right]^{\frac{\alpha}{3\alpha-2}}.\label{eq:FracMatLambda}
    \end{equation}
This scenario constitutes a minimal extension of the $\Lambda$CDM model and arises from the fractional cosmology modifications. We will consider it for our statistical analysis since the case $\alpha =2$ represents the concordance model.
    \item \textbf{Inclusion of Radiation:} When considering the early universe, the addition of radiation ($w_{r}=1/3$) extends the relation to:
    \begin{equation}
        \frac{H(z)}{H_{0}}=\left[\Omega_{r0}(1+z)^{4}+\Omega_{m0}(1+z)^{3}+\Omega_{\Lambda}\right]^{\frac{\alpha}{3\alpha-2}}.
    \end{equation}

    \item \textbf{The General Relativity Limit:} In the limit where the fractional parameter $\alpha \to 2$, the standard Bekenstein-Hawking entropy is recovered. As a result, the fractional exponent $\frac{\alpha}{3\alpha-2}$ approaches $\frac{1}{2}$, and the equation reduces smoothly to the usual Einstein–Friedmann relation:
    \begin{equation}
        \frac{H(z)}{H_{0}}=\sqrt{\sum_{i}\Omega_{i0}(1+z)^{3(1+w_{i})}}.
    \end{equation}
\end{itemize}

\section{Statistical analysis}
\label{ST}

In this section, we show how the constraints on the free parameters of the model are obtained through a joint statistical analysis based on late-time cosmological observations. We consider three complementary and statistically independent probes of the background expansion: Cosmic Chronometers (CC), Type Ia Supernovae (SNe Ia), and Baryon Acoustic Oscillations (BAO) from the DESI DR2 survey. Together, these datasets allow us to robustly explore parameter degeneracies and assess the consistency of the model with observational data.

Under the assumption that the individual datasets are uncorrelated, the total chi-square function is defined as the sum of the corresponding contributions,
\begin{equation}
\chi^2_{\rm tot} = \chi^2_{\rm CC} + \chi^2_{\rm SN} + \chi^2_{\rm DESI}.
\end{equation}

\subsection*{Cosmic Chronometers (CC)}

The cosmic chronometer technique provides direct estimates of the Hubble expansion rate $H(z)$ by exploiting the differential age evolution of passively evolving galaxies. By relating the redshift variation of galaxy ages to the cosmic expansion, this method offers a direct probe of $H(z)$ that is largely insensitive to assumptions about the cosmological distance scale.

In this work, we use a compilation of $N_{\rm CC}$ measurements $\{z_i, H_{\rm obs}(z_i), \sigma_{H_i}\}$ collected from the literature \cite{Moresco:2020fbm,cao2018cosmological,farooq2013hubble}. The corresponding chi-square estimator is defined as
\begin{equation}
\chi^2_{\rm CC} =
\sum_{i=1}^{N_{\rm CC}}
\frac{\left[ H_{\rm th}(z_i;\mathbf{p}) - H_{\rm obs}(z_i) \right]^2}{\sigma_{H_i}^2},
\end{equation}
where $H_{\rm th}(z_i;\mathbf{p})$ denotes the theoretical prediction for the Hubble rate evaluated at redshift $z_i$ for a given set of model parameters $\mathbf{p}$.

\subsection*{Type Ia Supernovae (SNe Ia)}

Type Ia supernovae act as standardized candles and provide high-precision measurements of relative cosmological distances through the distance modulus,
\begin{equation}
\mu_{\rm obs} = m_B^{\rm corr} - M,
\end{equation}
where $m_B^{\rm corr}$ is the light-curve corrected apparent magnitude and $M$ is the absolute magnitude of the supernovae.

For a given cosmological model, the theoretical distance modulus is computed as
\begin{equation}
\mu_{\rm th}(z;\mathbf{p}) = 5 \log_{10}\!\left[\frac{D_L(z;\mathbf{p})}{\mathrm{Mpc}}\right] + 25,
\end{equation}
with the luminosity distance defined by
\begin{equation}
D_L(z;\mathbf{p}) = (1+z)\int_0^z \frac{c\,dz'}{H(z';\mathbf{p})}.
\end{equation}

Our analysis is based on the PantheonPlus sample \cite{Brownsberger:2021uue,Brout:2022vxf,Scolnic:2021amr}\footnote{Available at \url{https://github.com/PantheonPlusSH0ES}.}, which consists of 1657 SNe Ia with redshifts $z>0.01$. The corresponding chi-square estimator is given by
\begin{equation}
\chi^2_{\rm SN} =
(\boldsymbol{\mu}_{\rm obs} - \boldsymbol{\mu}_{\rm th})^{T}
\,\mathbf{C}^{-1}\,
(\boldsymbol{\mu}_{\rm obs} - \boldsymbol{\mu}_{\rm th}),
\end{equation}
where $\mathbf{C}$ denotes the full covariance matrix accounting for both statistical and systematic uncertainties.

Since supernova observations probe only relative distances, they do not provide direct information on the absolute scale of the expansion rate. Consequently, the parameters $H_0$ and $M$ enter the analysis through a fully degenerate combination when SNe Ia data are considered in isolation. This degeneracy is resolved only through the inclusion of additional datasets that directly constrain the expansion rate.

Baryon acoustic oscillations provide geometrical distance measurements through the imprint of the sound horizon scale in the late-time distribution of matter. We make use of the most recent BAO measurements from the DESI DR2 release \cite{DESI:2025zgx,DESI:2025zpo}\footnote{Available at \url{https://github.com/CobayaSampler/bao_data/}.}, which include constraints on both transverse and radial distance indicators.

The observational data vector is defined as
\begin{equation}
\mathbf{X}_{\rm obs} =
\left\{
\frac{D_M(z)}{r_d},\,
\frac{D_H(z)}{r_d},\,
\frac{D_V(z)}{r_d}
\right\}_{\rm obs},
\end{equation}
where the transverse comoving distance is given by
\begin{equation}
    D_M(z) = \int_0^{z} \frac{c}{H(z')}\,dz',
\end{equation}
the Hubble distance by 
\begin{equation}
    D_H(z)=c/H(z),
\end{equation}
and the volume-averaged distance by
\begin{equation}
D_V(z) = \left[ D_M^2(z)\,\frac{c\,z}{H(z)} \right]^{1/3}.
\end{equation}
The BAO contribution to the likelihood is quantified through the chi-square function
\begin{equation}
\chi^2_{\rm DESI} =
(\mathbf{X}_{\rm obs} - \mathbf{X}_{\rm th})^{T}
\,\mathbf{C}^{-1}\,
(\mathbf{X}_{\rm obs} - \mathbf{X}_{\rm th}),
\end{equation}
where $\mathbf{C}$ is the covariance matrix provided by the DESI collaboration.\\

As the model considered in this work primarily affects the late-time cosmological dynamics, any modification of the early-Universe physics is expected to be subdominant. In particular, the impact on the sound horizon at the drag epoch is assumed to be negligible within the current observational precision. Under this assumption, we fix the sound horizon to its $\Lambda$CDM value, $r_d = 147.09\,\mathrm{Mpc},$ as determined by Planck 2018 observations \cite{Planck:2018vyg}.

\begin{figure}[!t]
\centering
\includegraphics[width=0.6\textwidth]{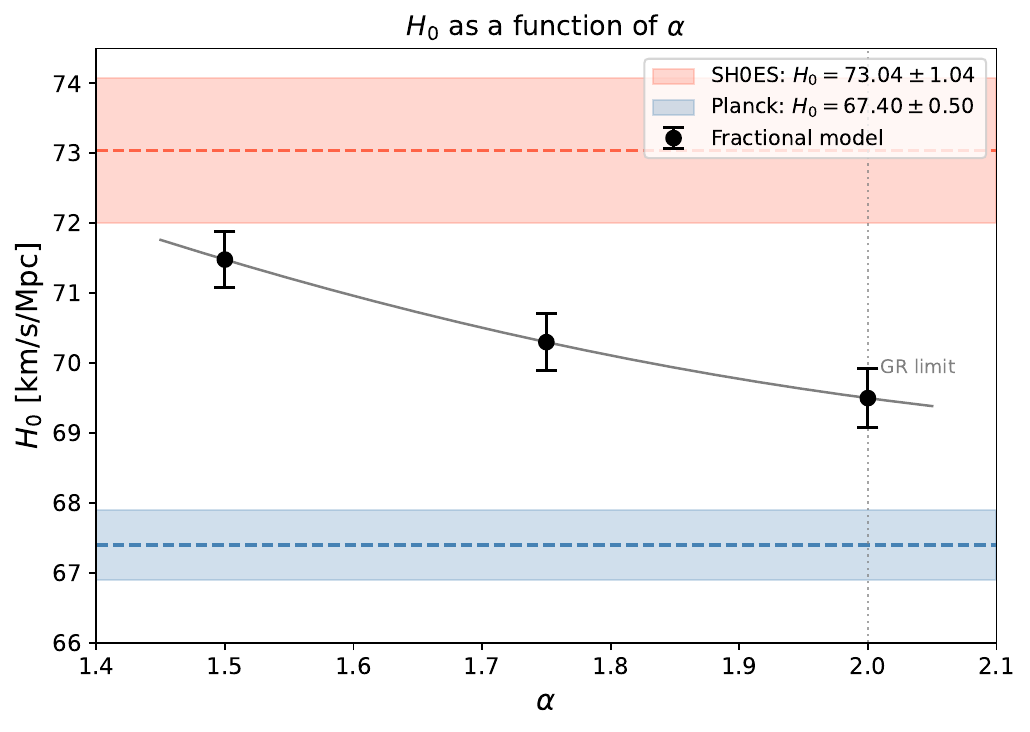}
\caption{Inferred values of $H_0$ as a function of the fractional 
parameter $\alpha$, obtained from the joint analysis of CC + 
PantheonPlus + SH0ES + DESI DR2 data. Error bars correspond to 
$1\sigma$ confidence level. The shaded bands indicate the 
$1\sigma$ regions for the SH0ES local measurement 
($H_0 = 73.04 \pm 1.04$ km/s/Mpc) and the Planck 2018 
inference ($H_0 = 67.40 \pm 0.50$ km/s/Mpc). The vertical 
dotted line marks the General Relativity limit $\alpha = 2$.}
\label{fig:H0_alpha}
\end{figure}

\begin{figure}[!h]
    \centering
    \includegraphics[width=0.6\linewidth]{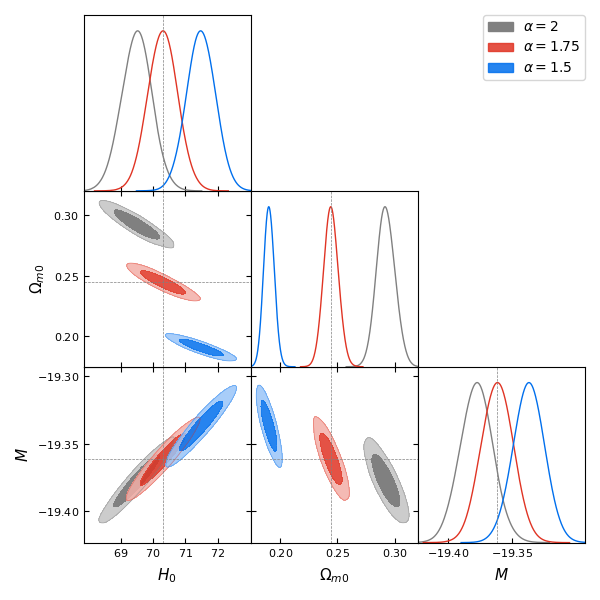}
    \caption{
    Marginalized posterior distributions and two-dimensional confidence contours
    ($68\%$ and $95\%$ CL) for the parameters
    $H_0$, $\Omega_{m0}$, and $M$, for fixed values of $\alpha$,
    obtained from the combined analysis of
    CC + PantheonPlus + SH0ES + DESI DR2 data.
    The dashed lines indicate the mean values of each parameter.
    }
    \label{fig:triangle_4params}
\end{figure}

\begin{figure}[!h]
    \centering
    \includegraphics[width=0.6\textwidth]{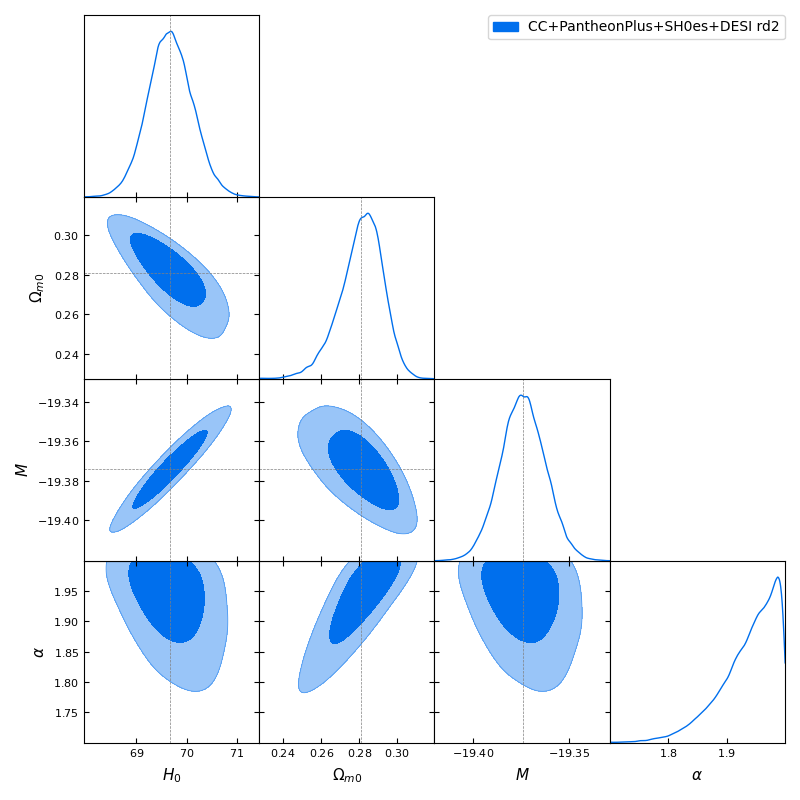}
    \caption{Triangle plot showing the marginalized posterior distributions and $68\%$ and $95\%$ confidence regions for the parameters $H_0$, $\Omega_{m0}$, $M$, and $\alpha$, obtained from the combined analysis of CC + PantheonPlus + SH0ES + DESI DR2. The posterior of $\alpha$ accumulates near the physical upper bound $\alpha=2$, indicating that the data favor the regime in which the model approaches its limiting case. Strong degeneracies are observed between $H_0$ and $\Omega_{m0}$, as well as between $\Omega_{m0}$ and $M$, while $\alpha$ shows clear correlations with the background parameters, reflecting its nontrivial role in shaping the late-time cosmological dynamics.}
    \label{fig:triangle_4param_alpha}
\end{figure}
\begin{figure}[b!]
\centering
\includegraphics[width=0.6\textwidth]{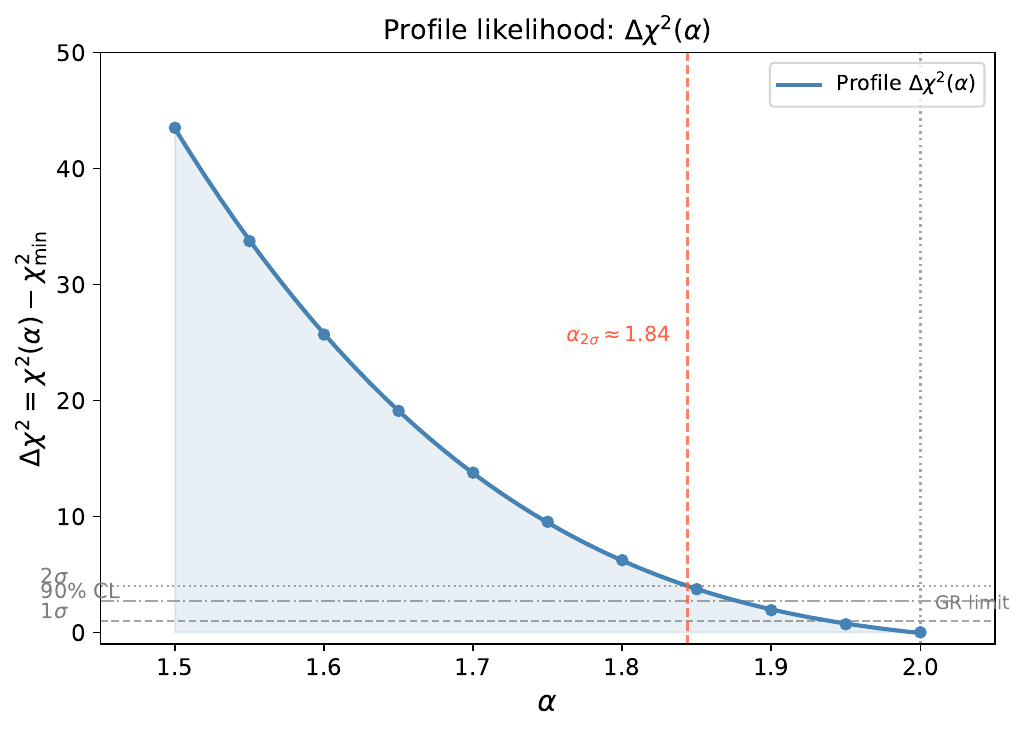}
\caption{Profile likelihood $\Delta\chi^2(\alpha) = 
\chi^2(\alpha) - \chi^2_{\min}$ as a function of the 
fractional parameter $\alpha$, with $H_0$, $\Omega_{m0}$, 
and $M$ profiled at each fixed value. Horizontal dashed 
lines indicate the $1\sigma$ ($\Delta\chi^2 = 1$), 90\% CL 
($\Delta\chi^2 = 2.71$), and $2\sigma$ ($\Delta\chi^2 = 4$) 
thresholds. The vertical dotted line marks the GR limit 
$\alpha = 2$.}
\label{fig:chi2}
\end{figure}

Figure~ \ref{fig:H0_alpha} and \ref{fig:triangle_4params} and table \ref{tab:constraints_model} present the constraints obtained from the combined analysis of CC + PantheonPlus + SH0ES + DESI DR2 data, considering fixed representative values of the parameter $\alpha$ within the physically motivated range $1 < \alpha \leq 2$. In this approach, we explore the conditional posterior distributions of $H_0$, $\Omega_{m0}$, and $M$ for $\alpha = 2, 1.75, 1.5$, enabling a direct assessment of the role of $\alpha$ in the background cosmology. However, we restrict the analysis to $\alpha \geq 1.5$, since lower values lead to best-fit solutions with unrealistically low values of $\Omega_{m0}$, which are phenomenologically disfavored. \\
For all considered values of $\alpha$, the posterior distributions of the remaining parameters remain well-defined and approximately Gaussian, indicating that the model is robustly constrained by late-time observations at the background level. A clear and consistent shift in the inferred cosmological parameters is observed as $\alpha$ varies. In particular, the joint contours reveal a strong anticorrelation between $H_0$ and $\Omega_{m0}$, consistent with expectations from late-time cosmological analyses. More importantly, varying $\alpha$ induces a coherent displacement along this degeneracy direction: decreasing $\alpha$ leads to higher values of $H_0$ and lower values of $\Omega_{m0}$. This behavior indicates that $\alpha$ effectively modulates the balance between matter content and the expansion rate, acting as a physically relevant degree of freedom. \\
A similar degeneracy is observed between $\Omega_{m0}$ and the supernova absolute magnitude $M$, reflecting their combined role in determining luminosity distances. This degeneracy is preserved across all fixed values of $\alpha$, although its location in parameter space shifts consistently with the variation of $\alpha$, as illustrated in the figures. \\
We now turn to the full parameter analysis, allowing $\alpha$ to vary within its physical domain, these results are shown in Figure \ref{fig:triangle_4param_alpha}. Using the combined dataset CC + PantheonPlus + SH0ES + DESI DR2, we obtain
\[
H_0 = 69.67^{+0.44 +0.88}_{-0.43 -0.86}, \quad 
\Omega_{m0} = 0.28^{+0.01 +0.02}_{-0.01 -0.02}, \quad 
M = -19.37^{+0.01 +0.02}_{-0.02 -0.03}, \quad 
\alpha = 1.94^{+0.06 +0.06}_{-0.02 -0.10},
\]
where the first (second) uncertainties correspond to the 68\% (95\%) confidence levels. From the goodness-of-fit perspective, the results exhibit a clear tendency toward larger values of $\alpha$. Since $\alpha = 2$ corresponds to the upper limit allowed by the physical domain of the model, the posterior distribution accumulates near this boundary, indicating that the observational data favor the regime in which the model approaches its limiting case. Within the physically allowed interval $1 < \alpha \leq 2$, no well-defined maximum is found in the interior of the parameter space. As a result, the analysis does not provide a precise determination of $\alpha$, but instead yields a robust lower bound,
\[
1.92 \leq \alpha \leq 2.00 \quad (68\% \, \text{CL}), \qquad
1.84 \leq \alpha \leq 2.00 \quad (95\% \, \text{CL}).
\]
Table~II extends this scan by reporting best-fit parameters 
across the full range $1.5 \leq \alpha \leq 2.00$ in steps 
of $\Delta\alpha = 0.05$. The results confirm the monotonic 
degradation of the fit quality as $\alpha$ decreases from 
the GR limit, with $\chi^2_\nu$ increasing from $0.886$ 
at $\alpha = 2$ to $0.911$ at $\alpha = 1.5$. This 
systematic behavior is further quantified in 
Fig.~\ref{fig:chi2}, which shows the profile likelihood 
$\Delta\chi^2(\alpha) = \chi^2(\alpha) - \chi^2_{\min}$, 
placing a lower bound of $\alpha \gtrsim 1.94$ at $1\sigma$ 
and $\alpha \gtrsim 1.84$ at $2\sigma$.
These constraints confirm that the data strongly prefer values 
of $\alpha$ close to the General Relativity limit while 
definitively excluding significant deviations from the 
standard Bekenstein–Hawking entropy at the background level. Furthermore, this consistency between the MCMC posterior and the profile likelihood strengthens the robustness of the inferred constraints on $\alpha$.

\begin{table}[t!]
\centering
\caption{Constraints obtained for fixed values of $\alpha$. We report mean values with $68\%$ confidence level uncertainties and the corresponding goodness-of-fit statistics.}
\label{tab:constraints_model}
\begin{tabular}{c|ccc}
\hline\hline
Parameter & $\alpha = 2$ & $\alpha = 1.75$ & $\alpha = 1.5$ \\
\hline\hline
$H_0$ & $69.50 \pm 0.42$ & $70.30 \pm 0.41$ & $71.48 \pm 0.40$ \\
$\Omega_{m0}$ & $0.292 \pm 0.008$ & $0.244 \pm 0.006$ & $0.190 \pm 0.004$ \\
$M$ & $-19.378 \pm 0.012$ & $-19.361 \pm 0.011$ & $-19.337 \pm 0.011$ \\
\hline
\multicolumn{4}{c}{\textit{Goodness of fit}} \\
\hline
$\chi^2_{\min}$ & $1502.99$ & $1512.53$ & $1546.50$ \\
$\chi^2_{\nu}$ & $0.886$ & $0.891$ & $0.911$ \\
\hline\hline
\end{tabular}
\end{table}

\begin{table}[t!]
\centering
\caption{Best-fit values and goodness-of-fit statistics for different values of $\alpha$.}
\label{tab:best_fit}
\begin{tabular}{c c c c c c}
\hline\hline
$\alpha$ & $H_0$ & $\Omega_{m0}$ & $M$ & $\chi^2_{\min}$ & $\chi^2_\nu$ \\
\hline\hline
1.50 & 71.481 & 0.190 & -19.337 & 1546.501 & 0.911 \\
1.55 & 71.213 & 0.201 & -19.343 & 1536.729 & 0.906 \\
1.60 & 70.955 & 0.212 & -19.348 & 1528.672 & 0.901 \\
1.65 & 70.724 & 0.223 & -19.353 & 1522.089 & 0.897 \\
1.70 & 70.500 & 0.234 & -19.357 & 1516.765 & 0.894 \\
1.75 & 70.295 & 0.245 & -19.362 & 1512.528 & 0.891 \\
1.80 & 70.123 & 0.254 & -19.365 & 1509.216 & 0.889 \\
1.85 & 69.962 & 0.264 & -19.368 & 1506.715 & 0.888 \\
1.90 & 69.781 & 0.274 & -19.372 & 1504.903 & 0.887 \\
1.95 & 69.645 & 0.283 & -19.375 & 1503.690 & 0.886 \\
2.00 & 69.503 & 0.292 & -19.377 & 1502.995 & 0.886 \\
\hline\hline
\end{tabular}
\end{table}

\subsection{Impact of the fractional parameter on background cosmological observables}

\begin{figure}[H]
\centering
\includegraphics[width=0.6\textwidth]{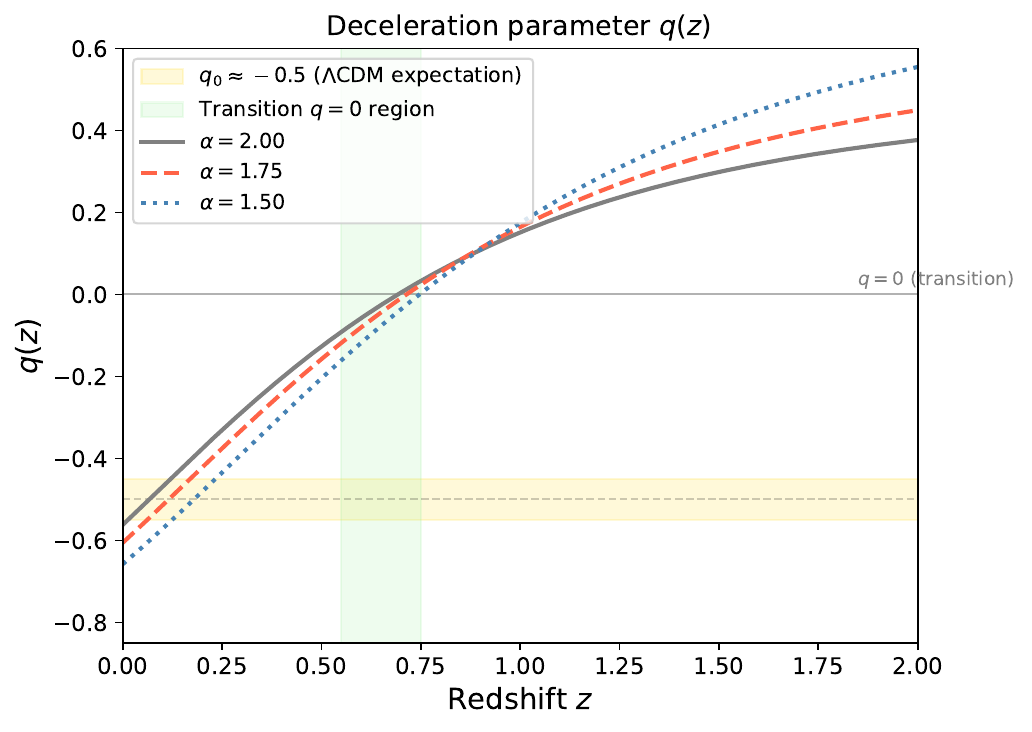}
\caption{Deceleration parameter $q(z)$ as a function of redshift 
for representative values of the fractional parameter $\alpha$, 
evaluated at the best-fit cosmological parameters of Table~I. 
The shaded green band indicates the approximate redshift interval 
of the deceleration-to-acceleration transition ($q = 0$). The 
gold band marks the expected present-day value $q_0 \approx -0.5$ 
for the concordance $\Lambda$CDM model. All curves recover the 
matter-dominated behavior $q \to 0.5$ at high redshift.}
\label{fig:qz}
\end{figure}

\begin{figure}[htbp!]
  \centering
  \includegraphics[width=0.6\linewidth]{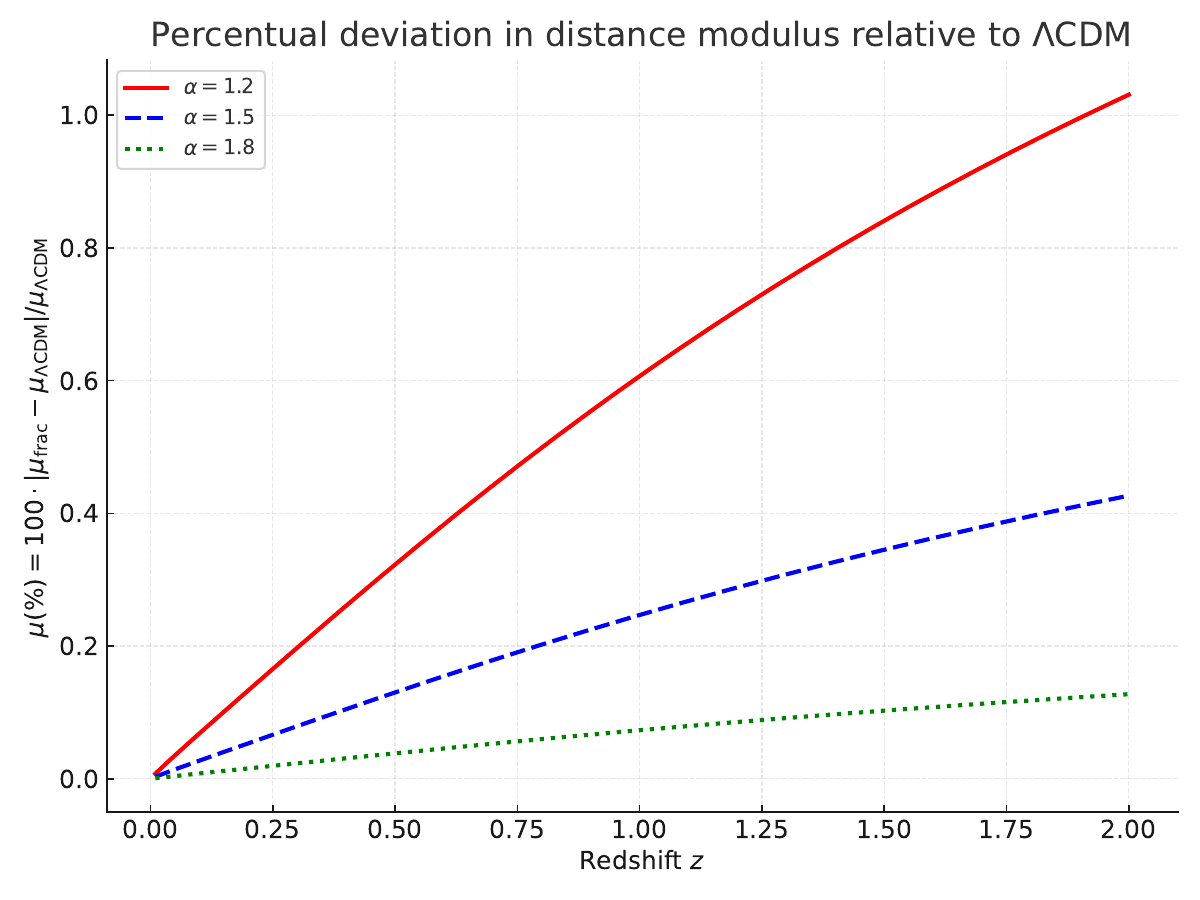}
  \caption{Percentage deviation of the distance modulus relative to the flat
  $\Lambda$CDM prediction.}
  \label{fig:mu_percent_error}
\end{figure}

The dynamical implications of the fractional parameter are 
further illustrated in Fig.~\ref{fig:qz}, which shows the 
deceleration parameter $q(z)$ for the three representative 
values of $\alpha$. All cases exhibit the expected 
transition from decelerated ($q > 0$) to accelerated 
($q < 0$) expansion. At $\alpha = 2$ the present-day value 
$q_0 \approx -0.56$ is consistent with the $\Lambda$CDM 
expectation, while decreasing $\alpha$ produces a more 
negative $q_0$ and shifts the transition redshift $z_t$ 
toward lower values, reflecting the enhanced effective 
dark energy contribution encoded in the fractional 
parameter.
On the other hand, in Fig. \ref{fig:mu_percent_error} we show the deviations between the fractional scenario and the concordance model through the relative percentage difference of the distance modulus, defined as
\begin{equation}
\mu(\%) = 100 \times 
  \frac{|\mu_{\rm frac}(z)-\mu_{\Lambda{\rm CDM}}(z)|}{\mu_{\Lambda{\rm CDM}}(z)},
\end{equation}
where $\mu_{\rm frac}(z)$ represents the fractional model specified by the parameter $\alpha$, and $\mu_{\Lambda\rm CDM}(z)$ stands for the prediction of the standard cosmological model. This quantity expresses, in percentage terms, how much the fractional model departs from the concordance scenario over the chosen redshift interval. Note that at the present epoch, both models overlap for different values of $\alpha$, showing that the fractional framework successfully reproduces the concordance model’s behavior at this stage. Introducing the additional parameter $\alpha$ produces deviations at the percent level in the distance modulus relative to the concordance model, indicating that the fractional scenario remains largely compatible with standard cosmology over part of cosmic history, while still allowing for potentially significant discrepancies in high-precision analyses. This also underlines how sensitive luminosity distance measurements are to even small deviations from the standard horizon entropy.
\section{Discussion and Conclusions}
In this work, we have systematically explored the cosmological and thermodynamic consequences of replacing the standard Bekenstein-Hawking area law with a fractional entropy functional $S_\alpha$ on the apparent horizon of an FLRW universe. Through the application of the unified first law of thermodynamics and the full Kodama-Hayward surface temperature, we derived a modified set of Friedmann equations. These generalized evolution equations demonstrate that modifying the microstate statistics of the horizon inherently alters the macroscopic gravitational dynamics, driving the late-time accelerated expansion of the universe.

From a purely thermodynamic perspective, we evaluated the stability of this fractional cosmology by calculating the specific heats at constant volume and constant pressure. Unlike other modified gravity scenarios or extended entropy models that exhibit horizon phase transitions, our model yields specific heats $C_V$ and $C_p$ that share the same sign and evolve strictly with the deceleration parameter $q(t)$. Cosmologically, the absence of divergences or sign-switching in the specific heats is highly significant: it confirms that the universe governed by fractional entropy avoids the catastrophic thermodynamic instabilities often found in alternative dark energy models. This guarantees a smooth, continuous cosmic expansion history that mimics the stabilizing role of the cosmological constant, ensuring the model remains thermodynamically well-behaved throughout the transition into the dark energy-dominated epoch.

At the background level, the viability of the truncated fractional cosmology was tested using a robust joint analysis of independent late-time probes: Cosmic Chronometers, Type Ia Supernovae from the Pantheon+ sample, the SH0ES prior, and the most recent BAO measurements from the DESI DR2 survey. First, we explored its impact by considering representative values within the physically motivated range $1 < \alpha \leq 2$. Within this framework, the model provides stable and well-constrained estimates for the cosmological parameters, with unimodal posterior distributions in all cases. Within this range, the fractional framework remains fully compatible with current observational constraints, achieving a level of precision comparable to the concordance $\Lambda$CDM model ($\chi^2_\nu \approx 0.886$), while effectively mimicking dark energy through the geometric deformation encoded in $\alpha$. We then performed a full MCMC analysis allowing all background parameters, including $\alpha$, to vary simultaneously. Despite the presence of the physical upper bound $\alpha = 2$, corresponding to the GR limit, the statistical analysis still enables the extraction of robust constraints on the model. In particular, the posterior distribution accumulates near the boundary but remains sufficiently informative to establish a well-defined lower limit, yielding $\alpha \gtrsim 1.92$ at $1\sigma$ and $\alpha \gtrsim 1.84$ at $2\sigma$. This demonstrates that the theory can be meaningfully constrained by observational data even in the presence of a bounded parameter space.

Our observational constraints on the fractional parameter are in excellent agreement with recent independent studies exploring Fractional Holographic Dark Energy (FHDE). Parallel investigations have successfully reconstructed FHDE using scalar and gauge fields \cite{Bidlan2025}, and tested its observational viability against Cosmic Chronometers, BAO, and supernova datasets, confirming the robustness of fractional models in driving late-time acceleration \cite{Trivedi2024, Rathore2025, Mazumdar2026}. Furthermore, just as our model preserves thermodynamic stability, FHDE models have been shown to offer novel perspectives on avoiding or classifying future singular scenarios, such as the Little Rip or Pseudo-Rip \cite{Bidlan2026}. This collective body of evidence underscores that fractional entropy is a highly competitive and observationally testable framework for dark energy.

While the background dynamics is heavily constrained by current probes, the fractional degrees of freedom may still leave distinct imprints on the formation of large-scale structures. Future extensions of this work should analyze the evolution of cosmological perturbations within the fractional framework. Specifically, deriving the effective Newtonian gravitational coupling, $G_{\mathrm{eff}}(a, k)$, and testing it against high-precision Redshift-Space Distortion (RSD) and weak lensing data will be critical. Such an analysis could break the existing degeneracies between $\alpha$ and $\Omega_{m0}$, opening new observational windows to conclusively distinguish fractional entropy cosmologies from the standard $\Lambda$CDM paradigm.

\section*{Acknowledgments}
M.~Cruz work was partially supported by S.N.I.I. (SECIHTI-M\'exico). S.~Lepe acknowledges the FONDECYT grant N°1250969, Chile. J.~Saavedra acknowledges the FONDECYT grant N°1220065, Chile.

\section*{Data Availability Statement}
No new datasets were generated or analysed during the current study.


\end{document}